\documentclass[aps,prb,preprint,superscriptaddress]{revtex4-1}
\usepackage{amssymb}
\usepackage[utf8]{inputenc}
\usepackage{graphicx}
\usepackage{bm,color,subfigure,amsmath}

\begin{document}

\title{Vortex-antivortex proliferation from an obstacle in thin film ferromagnets}

\author{Ezio~Iacocca}
\email{ezio.iacocca@colorado.edu}
\email{ezio@chalmers.se}
\affiliation{Department of Applied Mathematics, University of Colorado, Boulder, Colorado 80309-0526, USA}
\affiliation{Department of Physics, Division for Theoretical Physics, Chalmers University of Technology, 412 96, Gothenburg, Sweden}

\author{Mark~A.~Hoefer}
\affiliation{Department of Applied Mathematics, University of Colorado, Boulder, Colorado 80309-0526, USA}

\begin{abstract}
  Magnetization dynamics in thin film ferromagnets can be studied
  using a dispersive hydrodynamic formulation. The equations
  describing the magnetodynamics map to a compressible fluid with
  broken Galilean invariance parametrized by the longitudinal spin
  density and a magnetic analog of the fluid velocity that define
  spin-density waves. A direct consequence of these equations is the
  determination of a magnetic Mach number. Micromagnetic simulations
  reveal nucleation of nonlinear structures from an impenetrable
  object realized by an applied magnetic field spot or a defect. In this
  work, micromagnetic simulations demonstrate vortex-antivortex pair
  nucleation from an obstacle.  Their interaction establishes either
  ordered or irregular vortex-antivortex complexes. Furthermore, when
  the magnetic Mach number exceeds unity (supersonic flow), a Mach
  cone and periodic wavefronts are observed, which can be well-described
  by solutions of the steady, linearized equations. These results
  are reminiscent of theoretical and experimental observations in
  Bose-Einstein condensates, and further supports the analogy between
  the magnetodynamics of a thin film ferromagnet and compressible
  fluids. The nucleation of nonlinear structures and vortex-antivortex
  complexes using this approach enables the study of their
  interactions and effects on the stability of spin-density waves.
\end{abstract}

\maketitle

\section{Introduction}

Magnetization dynamics in thin film ferromagnets provide an exciting
platform to study nonlinear wave phenomena. This is possible due to
the exchange interaction that confers ferromagnetic order and wave
dispersion, nonlinear effects such as anisotropy, and diverse
mechanisms available to excite magnetization
dynamics~\cite{Stor2006}. In fact, coherent nonlinear magnetization
structures were observed many decades ago, such as domain walls and
vortices~\cite{Hubert2009}. More recently, envelope
solitons~\cite{Tong2010,Wang2014}, dissipative
droplets~\cite{Hoefer2010,Mohseni2013,Iacocca2014,Macia2014,Chung2016},
and skyrmions~\cite{Romming2013,Sampaio2013,Zhou2015,Jiang2015} have
also been observed in these materials. Another class of nonlinear
structure in thin film ferromagnets is a spatially periodic,
superfluid-like texture or soliton
lattice~\cite{Konig2001,Sonin2010,Takei2014,Chen2014,Iacocca2016b}
that is able to carry spin currents due to its nontrivial topology,
opening new pathways to nonlinear phenomena and potential
applications.

Recently, superfluid-like magnetic states have been interpreted in the context of
hydrodynamics~\cite{Halperin1969,Iacocca2016b}. It was shown that the
Landau-Lifshitz (LL) equation describing magnetodynamics can be
exactly cast as dispersive hydrodynamic equations, reminiscent of
those describing Bose-Einstein condensates (BECs)~\cite{Pethick2002}
and other superfluid-like media~\cite{el_dispersive_2016}. An additional, direct, exact connection between polarization waves in two-component
BECs~\cite{Stamper2013} and magnetization dynamics has recently been
identified~\cite{Qu2016,Congy2016}. From the magnetic dispersive
hydrodynamic formulation, it is possible to characterize superfluid-like states as dynamic, uniform hydrodynamic states (UHSs) or static,
spin-density waves (SDWs) parametrized by a longitudinal spin density
$n$ and fluid velocity $\mathbf{u}$ proportional to the spatial
gradient of the magnetization's in-plane phase. By defining the
magnetic analog of the Mach number from classical gas dynamics,
$\mathrm{M}$, subsonic and supersonic flow conditions can be
identified. Interestingly, this magnetic fluid representation, due to
ferromagnetic exchange coupling, generally exhibits broken Galilean invariance
at the level of (linear) spin wave excitations~\cite{Iacocca2016b}.
Consequently, the physics are reference-frame-dependent. This is in
stark contrast to the velocity-dependent dynamics of localized,
topological textures that result from their inherent
nonlinearity~\cite{Yamada2007,Beach2008}.

Intriguing dynamics result from the interaction between a superfluid-like state and a finite-sized obstacle. Micromagnetic
simulations~\cite{Iacocca2016b} demonstrated that, in general, SDWs
($\mathbf{u}\neq0$) at subsonic conditions, $\mathrm{M}<1$, flow in a
stable, laminar fashion around a point defect whereas a Mach cone,
wavefronts, and irregular vortex-antivortex (V-AV) pair nucleation
takes place at supersonic conditions, $\mathrm{M}>1$.  In the
different regime of thick, homogeneous ferromagnets ($\mathbf{u} =
0$), ``spin-Cerenkov'' radiation was observed in the moving reference
frame~\cite{Yan2013}. In classical fluids, similar coherent structures
can be nucleated from an obstacle. The conditions defining the onset
of specific features usually depend on the flow velocity or,
equivalently, the Mach number. At subsonic conditions, obstacles can
nucleate vortices resulting from an unsteady
wake~\cite{Williamson1996,Leweke2016}. A common example is the von
K\'{a}rm\'{a}n vortex street, characterized by a train of vortices of
alternating circulation. This has been thoroughly studied as a
function of the Reynolds number for viscous, incompressible
fluids~\cite{Williamson1996}. At supersonic conditions for a
compressible gas, a Mach cone can be generated as typified by a jet
breaking the speed of sound.

In contrast, superfluids such as BECs exhibit important
differences. Firstly, vortices are quantized due to irrotational flow
and secondly, BECs are compressible and inviscid. Therefore, a
Reynolds number is not defined in the sense of classical
fluids. However, it has been suggested that an alternative superfluid
Reynolds number can be defined from the onset of quantum vortex
shedding~\cite{Reeves2015}. In fact, V-AV pair dynamics can occur in
superfluids as their numerical nucleation~\cite{Frisch1992} and
instability~\cite{Nore1993} were demonstrated at subsonic
conditions. Numerical studies also identified the existence of a von
K\'{a}rm\'{a}n like vortex street~\cite{Sasaki2010}, which has been
recently demonstrated experimentally~\cite{Kwon2016}. At supersonic
conditions, Mach cones have been observed, both theoretically and
experimentally, accompanied by wave radiation or wavefronts ahead of
the obstacle~\cite{Carusotto2006,Gladush2007} and, at large
$\mathrm{M}$, steady, oblique dark solitons can be generated inside
the Mach cone~\cite{El2006}.

The similarity between the equations describing BECs and thin film
ferromagnetic magnetodynamics suggests that the latter may support the
many structures mentioned above, with the possibility of new phenomena
due to a ferromagnet's more complex geometry. The stability of SDWs
in a finite, thin film ferromagnet may be impacted by V-AV
nucleation. It is known that V-AV pairs in magnets can excite spin
waves by diverse annihilation processes~\cite{Hertel2006}, e.g., V-AV
interaction with other V-AVs, with defects, or with physical
boundaries. To explore the existence, dynamics, and stability of
nonlinear structures in thin film ferromagnets, we study analytically and
numerically the interaction between a SDW and an impenetrable,
finite-sized obstacle or defect.

In this paper, we show that in the static, laboratory reference frame,
trains of V-AV pairs nucleated by a sufficiently large obstacle
exhibit stable, linear motion at subsonic conditions whereas irregular
V-AV dynamics are observed at supersonic conditions. This is due to
the V-AV translation imposed by the underlying fluid velocity and the
interactions between vortices, leading to translational instabilities
or even V-V and AV-AV rotation. In the moving reference frame with
zero fluid velocity, V-AV pairs generally annihilate, leading to
irregular dynamics. However, at supersonic conditions, V-AV pairs
exhibit structure by describing an oblique path inside the Mach
cone. Wavefronts are observed to nucleate ahead of the obstacle in
both reference frames at supersonic conditions. For thin film
ferromagnets with finite extent, the observed nonlinear structures are
transient as the system relaxes to an energy minimum. These results
are not only relevant for the stability of a SDW but they also represent a
cornerstone to study V-AV complexes and their interaction with other
nonlinear structures. More generally, our work provides an avenue to
study the proliferation of topological textures in magnetism. For example,
recent numerical results have demonstrated the generation of skyrmions
by spin-transfer torque from anisotropic
obstacles~\cite{Everschor2016}, similar to the linear V-AV vortex
motion we show below.

The paper is organized as follows. Section II summarizes the
hydrodynamic formulation of the Landau-Lifshitz equation. The magnetic
dispersive hydrodynamic equations obtained in section II are compared
to the equations describing the mean field dynamics of BECs and
two-component BECs in section III. In section IV, we use a linearized
analysis to predict some properties of the patterns supported by the
dispersive hydrodynamic flow past an obstacle. We also use analogies
to classical fluids and superfluids to identify common flow patterns. Section
V describes the results obtained from micromagnetic
simulations. Finally, we provide our concluding remarks in section VI.

\section{Dispersive hydrodynamic formulation}

Following the formulation outlined in Ref.~\onlinecite{Iacocca2016b},
the magnetization dynamics of a thin film ferromagnet with planar
anisotropy can be conveniently described by the nondimensionalized LL
equation
\begin{equation}
\label{eq:1}
\frac{\partial\mathbf{m}}{\partial t} =
-\mathbf{m}\times\mathbf{h}_\mathrm{eff}-\alpha\mathbf{m}\times\mathbf{m}
\times\mathbf{h}_\mathrm{eff}, 
\end{equation}
with an effective field given by
\begin{equation}
\label{eq:2}
  \mathbf{h}_\mathrm{eff} = \Delta\mathbf{m}-m_z\hat{\mathbf{z}}+h_0\hat{\mathbf{z}},
\end{equation}
including, respectively, the ferromagnetic exchange field, a local (zero-thickness limit) dipolar field, and a perpendicular external field. The dispersive hydrodynamic equations are obtained by inserting the transformation to Hamiltonian variables~\cite{papanicolaou_dynamics_1991}
\begin{equation}
  \label{eq:3}
  n = m_z,\quad
  \mathbf{u} = - \nabla \Phi = -\nabla \left [
    \arctan{\left(m_y/m_x\right)} \right ],
\end{equation}
into Eq.~\eqref{eq:1}, where $|n|\le 1$ is the longitudinal spin
density ($|n| = 1$ corresponds to the vacuum state) and $\mathbf{u}$
is the curl-free (irrotational) fluid velocity. First, we exactly solve for
$\partial\Phi/\partial t$ and obtain
\begin{eqnarray}
  \label{eq:4}
    \frac{\partial \Phi}{\partial t} &=& -(1 - | \mathbf{u} |^2) n  + \frac{\Delta n}{1 - n^2} + \frac{ n |\nabla n|^2}{(1-n^2)^2}\nonumber\\
    &&+h_0- \frac{\alpha}{1 - n^2} \nabla \cdot [(1-n^2) \mathbf{u}].
\end{eqnarray}

The gradient of Eq.~\eqref{eq:4} and the equation for $n$ are
\begin{subequations}
\label{eq:5}
\begin{eqnarray}
  \label{eq:51}
    \frac{\partial n}{\partial t} &=&
    \nabla\cdot\left[(1-n^2)\mathbf{u}\right] +
    \alpha(1-n^2)\frac{\partial \Phi}{\partial
        t} ,\\
  \label{eq:52}
  \frac{\partial \mathbf{u}}{\partial t} &=&
  \nabla\left[(1-|\mathbf{u}|^2)n
    \right] 
  - \nabla\left[\frac{\Delta n}{1-n^2}+\frac{n |\nabla
        n|^2}{(1-n^2)^2}\right]\nonumber\\
				&&-\nabla h_0
  +\alpha\nabla\left[\frac{1}{1-n^2}\nabla\cdot\left[(1-n^2) 
        \mathbf{u}\right]\right].
\end{eqnarray}
\end{subequations}
These are an exact transformation of the LL equation,
Eq.~\eqref{eq:1}, and fully describe the magnetization dynamics.

The ground-state solutions of Eqs.~\eqref{eq:51} and \eqref{eq:52} are static magnetization textures known as spin-density waves (SDWs), such that 
\begin{equation}
\label{eq:53}
  \frac{\partial\Phi}{\partial t}=\Omega=-(1-\bar{u}^2)\bar{n}+h_0=0
\end{equation}
where the longitudinal spin density and fluid velocity,
$(n,\mathbf{u})=(\bar{n},\bar{u}\hat{\mathbf{x}})$, are
constant. This implies that the SDW can have a constant out-of-plane
tilt and a periodic, in-plane spatial rotation of the azimuthal angle
$\Phi$. For simplicity, we consider a non-negative fluid velocity
along $\hat{\mathbf{x}}$, i.e., $\bar{u}>0$.

The static condition Eq.~\eqref{eq:53} at a finite field corresponds
to the ground state SDW conditions imposed by magnetic damping. If by
some means $\Omega\neq0$ in Eq.~\eqref{eq:53}, e.g., by an abrupt jump
in the field $h_0$ to another value $h_1$, the associated dynamic
solutions are considered UHSs~\cite{Iacocca2016b} that undergo a slow relaxation to a SDW.
This relaxation process for $\bar{n}(t)$ maintains constant $\bar{u}$
and can be computed by assuming spatial uniformity in
Eqs.~\eqref{eq:51} resulting in the temporal differential equation for
the longitudinal spin density
\begin{equation}
  \label{eq:SM15}
  \frac{d\bar{n}}{d t} =
  -\alpha(1-\bar{u}^2)(1-\bar{n}^2)\left ( \bar{n} - 
   \frac{h_1}{1 - \bar{u}^2} \right ) .
\end{equation}
Upon integration, Eq.~\eqref{eq:SM15} yields the implicit relationship
\begin{equation}
  \label{eq:SM16} 
    \left [ \frac{\left ( \bar{n} - \frac{h_1}{1-\bar{u}^2} \right 
        )^2}{1-\bar{n}^2} \right ]^{1-\bar{u}^2} \left (
      \frac{1+\bar{n}}{1-\bar{n}} 
    \right )^{h_1} = C \exp \left \{ -2\alpha \left [ (1-\bar{u}^2)^2
        - h_1^2 \right ] t \right \} ,
\end{equation}
for $\bar{n}(t)$, where the constant $C$ is determined from the
initial density $\bar{n}_0 = h_0/(1-\bar{u}^2)$.  This expression
composes an exponential decay of the density to the static state
$\bar{n}(t) \to \bar{n}_1 = h_1/(1-\bar{u}^2)$, $t \to \infty$ for any
initial $\bar{u}$.  

To study the dynamics originating from the interaction between a SDW
and an obstacle, it is important to characterize small-amplitude
perturbations of the SDW, described by the generalized spin wave
dispersion relation
\begin{equation}
  \label{eq:6}
  \omega_\pm = \left(2 \bar{n}
    \mathbf{u}-\mathbf{V}\right)\cdot \mathbf{k} \pm 
  |\mathbf{k}|\sqrt{(1-\bar{n}^2) (1-\bar{u}^2)+|\mathbf{k}|^2}, 
\end{equation}
where $\mathbf{k}$ is the wave vector and $\mathbf{V}$ is the velocity
of an external observer or Doppler shift. This generalized dispersion
relation reduces to the typical Galilean invariant, exchange-mediated
spin wave dispersion in the vacuum limit $|\bar{n}|\approx 1$. However,
Galilean invariance is broken in general~\cite{Iacocca2016b}, leading
to reference-frame-dependent physics.

From Eq.~\eqref{eq:6}, one can derive the generalized spin wave phase and group velocities, respectively, $v_p$ and $v_g$
\begin{subequations}
\begin{eqnarray}
  \label{eq:71}
  \mathbf{v}_{p,\pm} &=&\frac{\omega_\pm}{|\mathbf{k}|}\hat{\mathbf{k}} \\
  &=&\left[\left(2\bar{n}\mathbf{u}-\mathbf{V}\right)\cdot\hat{\mathbf{k}}\pm
  \sqrt{(1-\bar{n}^2) 
    (1-\bar{u}^2)+|\mathbf{k}|^2}\right]\hat{\mathbf{k}},\nonumber\\  
  \label{eq:72}
  \mathbf{v}_{g,\pm} &=& \nabla_\mathbf{k}\omega_\pm\hat{\mathbf{k}} \\
	&=&
    \left[\left(2\bar{n}\mathbf{u}-\mathbf{V}\right)\cdot\hat{\mathbf{k}}\pm
    \frac{(1-\bar{n}^2)
    (1-\bar{u}^2)+2|\mathbf{k}|^2}{\sqrt{(1-\bar{u}^2)
      (1-\bar{n}^2)+|\mathbf{k}|^2}}\right]\hat{\mathbf{k}}.\nonumber  
\end{eqnarray}
\end{subequations}

The long-wavelength limit $|\mathbf{k}|\rightarrow0$ leads to
coincident phase and group velocities, $\mathbf{v}_p=\mathbf{v}_g$, corresponding to the
magnetic sound speeds imposed by the SDW
\begin{equation}
  \label{eq:8}
  s_\pm = 2\bar{n} \bar{u} + \bar{V} \pm \sqrt{(1 - \bar{n}^2)(1- \bar{u}^2)},
\end{equation}
where we have assumed $\mathbf{V}=-\bar{V}\hat{\mathbf{x}}$ and
$\hat{\mathbf{k}} = \hat{\mathbf{x}}$ for simplicity.

Subsonic (counter-propagating waves) and supersonic (co-propagating
waves) flow conditions can be identified from Eq.~\eqref{eq:8}. In
particular, the transition between these regimes, the sonic curve,
occurs when $s_- = 0$ or $s_+ = 0$ in Eq.~\eqref{eq:8}, resulting in
\begin{equation}
  \label{eq:9}
  \bar{V}\left(\bar{V}+4\bar{n}\bar{u}\right) =
  \left(1-\bar{n}^2\right)-\left(1+3\bar{n}^2\right)\bar{u}^2.
\end{equation}

\begin{figure}[t]
\centering \includegraphics[width=3.in]{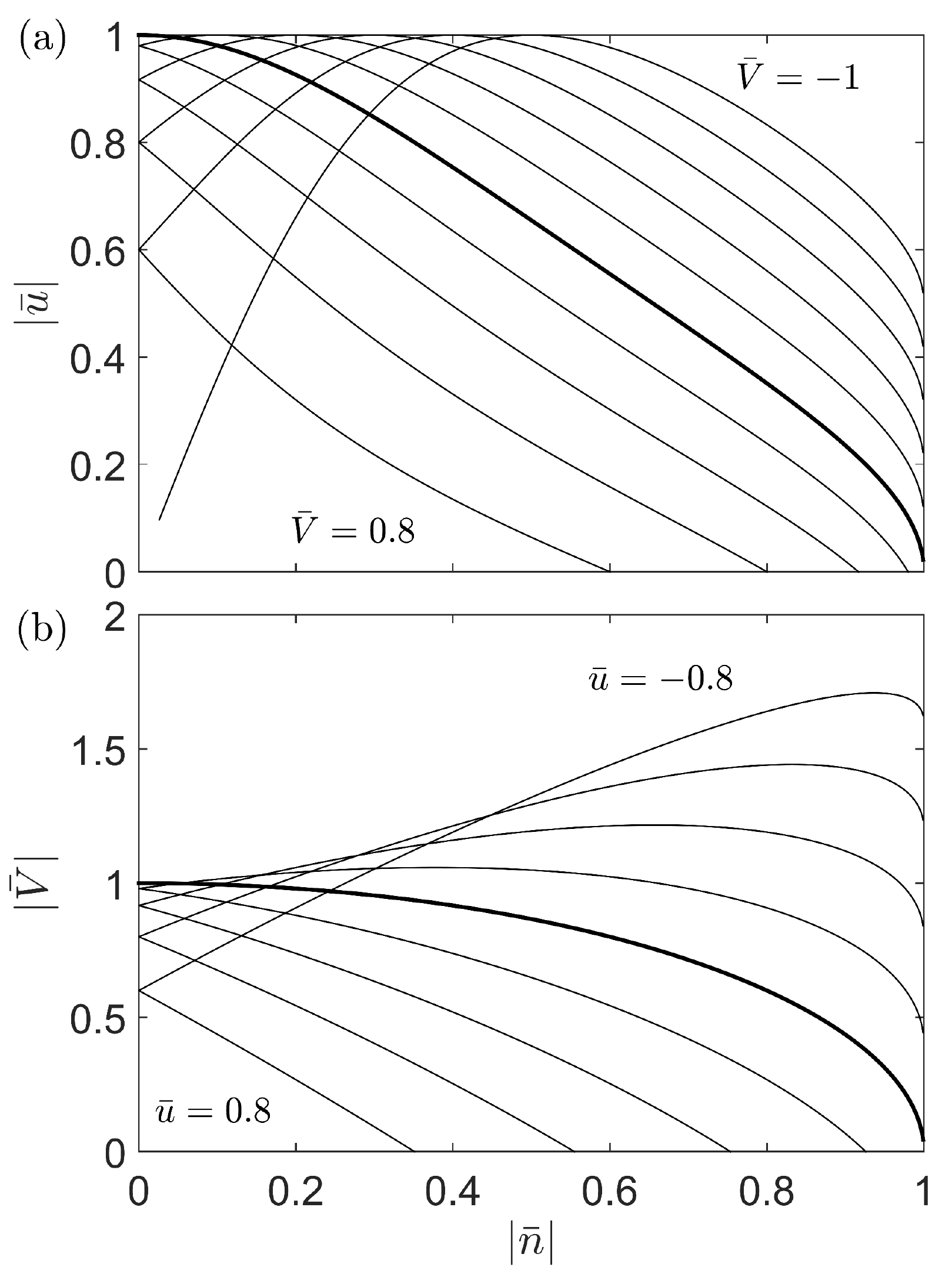}
\caption{ \label{fig1} Sonic curves calculated from Eq.~\eqref{eq:9} by varying (a) $\bar{V}$ and (b) $\bar{u}$ in steps of $0.2$. The cases for (a) $\bar{V}=0$ and (b) $\bar{u}=0$ are emphasized by the thicker solid line, corresponding to Eq.~\eqref{eq:10}. }
\end{figure}

Equation~\eqref{eq:9} represents the sonic surface for any $\bar{V}$, $\bar{u}$, and $|\bar{n}|<1$, projections of which are shown in Fig.~\ref{fig1}. Relatively simple expressions for Eq.~\eqref{eq:9} are available when we restrict to $\bar{V}=0$ or $\bar{u}=0$ (thicker lines in Fig.~\ref{fig1}).  Thus, we define the Mach numbers in these cases
\begin{equation}
  \label{eq:10}
  \mathrm{M}_{u} =
  |\bar{u}|\sqrt{\frac{1+3\bar{n}^2}{1-\bar{n}^2}}, \quad 
  \mathrm{M}_{V} = \frac{|\bar{V}|}{\sqrt{1-\bar{n}^2}},
\end{equation}
respectively, so that $\mathrm{M} = 1$ corresponds to Eq.~\eqref{eq:9}.

\section{Bose-Einstein condensate limit of the dispersive hydrodynamic
  equations}
\label{sec:bose-einst-cond}

The hydrodynamic equations governing the mean field dynamics of a BEC can be obtained as a limiting case of the magnetic, dispersive hydrodynamic equations, Eqs.~\eqref{eq:51} and \eqref{eq:52}.  For this, one should consider the nearly perpendicular, small velocity, long-wavelength, and low-frequency expansion
\begin{eqnarray}
  \label{eq:11}
    n &=& 1 - \epsilon \rho, \quad \mathbf{u} = \left ( \frac{1}{2}\epsilon
    \right )^{1/2}
    \mathbf{v},\nonumber\\ \tilde{\mathbf{x}} &=&
    \left ( \frac{1}{2}\epsilon \right )^{1/2} \mathbf{x}, \quad 
    \tilde{t} = \epsilon t, \quad U = \epsilon h_0,
\end{eqnarray}
where $0 < \epsilon \ll 1$ measures the magnetization deviation from
the perpendicular, vacuum state.  Inserting this expansion into
Eqs.~\eqref{eq:51} and \eqref{eq:52} while keeping only the leading
order terms and setting $\alpha=0$ results in
\begin{subequations}
\label{eq:12}
\begin{eqnarray}
  \label{eq:121}
  \frac{\partial \rho}{\partial \tilde{t}} &+& \tilde{\nabla} \cdot
  ( \rho \mathbf{v} ) = 0 , \\
  \label{eq:122}
  \frac{\partial \mathbf{v}}{\partial \tilde{t}} &+&
  (\mathbf{v}\cdot \tilde{\nabla}) \mathbf{v} + \tilde{\nabla}
  \rho = \frac{1}{4} \tilde{\nabla} \left [ \frac{\tilde{\Delta}
      \rho}{\rho} - \frac{|\tilde{\nabla}
      \rho|^2}{2\rho^2} \right ] - \tilde{\nabla} U,
\end{eqnarray}
\end{subequations}
a non-dimensional form of the conservative, hydrodynamic equations of
a repulsive BEC with trapping potential $U(\tilde{\mathbf{x}})$ (see,
e.g., Ref.~\onlinecite{Pethick2002}). From this analogy
it follows that the ``healing length'' defined in BEC as the
transition distance between two states with dissimilar density is
simply given by the spatial scaling in ferromagnetic materials, i.e.,
the exchange length in the case of planar
ferromagnets~\cite{Iacocca2016b}.

BECs can lose superfluidity when its interaction with an obstacle
results in the generation of small-amplitude
waves~\cite{Pethick2002}. This can be described in terms of the Landau
criterion for superfluidity, which invokes Galilean invariance to find
the conditions where spontaneous wave generation is energetically
unfavorable. For a BEC described by Eq.~\eqref{eq:12}, the Landau
criterion is $|\mathbf{v}| < s$ where $s = \sqrt{\rho}$ is the BEC
long wave speed of sound.

For BECs, the Landau criterion and the Mach number are closely
related. Both describe the conditions for subsonic to supersonic flow
transitions.  The Mach number for BECs is~\cite{Pethick2002}
\begin{equation}
\label{eq:13}
  \mathrm{M}_\mathrm{BEC} = \frac{|\mathbf{v}|}{\sqrt{\rho}} .
\end{equation}
The sonic curve, ${\mathrm{M}_\mathrm{BEC}=1}$, is the critical
transition from subsonic to supersonic conditions and represents the
breakdown of superfluidity, i.e., spontaneous wave generation is
energetically favorable. It can be verified that the magnetic Mach
numbers, Eq.~\eqref{eq:10}, reduce to Eq.~\ref{eq:13} for small
deviation from vacuum, by use of the transformation \eqref{eq:11}.  Since the standard derivation of the Landau
criterion utilizes Galilean invariance, it can \textit{only} be
applied to exchange-mediated, Galilean invariant spin waves in a
perpendicularly magnetized thin film ferromagnet.  Away from this
regime, SDWs break Galilean invariance~\cite{Iacocca2016b}.  The more
general identification of the subsonic to supersonic transition as the
breakdown of superfluidity, Eq.~\eqref{eq:10}, is the appropriate one for SDWs because the
generalized spin wave dispersion \eqref{eq:6} exhibits non-zero
curvature for positive wavenumbers.

It is important to stress that Eqs.~\eqref{eq:12} are conservative, so
that the connection between BECs and exchange-mediated spin waves is
valid insofar as magnetic damping is neglected. However, because
damping is typically weak $0 < \alpha \ll 1$, we can invoke
conservative arguments to describe the dynamics and nucleated
structures over sufficiently short timescales (proportional to
$\alpha^{-1}$) when a finite-sized obstacle is introduced.

A more general analog to the magnetic, dispersive hydrodynamic
equations, Eqs.~\eqref{eq:51} and \eqref{eq:52}, can be found in
two-component Bose-Einstein condensates, a class of spinor Bose
gases~\cite{Stamper2013} that possess magnetic properties. In
particular, Congy \emph{et al.}~\cite{Congy2016} found that the
polarization waves in a one-dimensional, two-component BEC can be
described by approximate dispersive hydrodynamic equations coinciding
exactly with the one-dimensional projection of Eqs.~\eqref{eq:51} and
\eqref{eq:52} when $\alpha = 0$. This suggests that observations in
planar ferromagnets can be applicable to two-component BECs and
showcases the effects of exchange coupling between spins or atoms with
a finite magnetic moment. Magnetic damping is an energy sink for a
planar ferromagnet, at which point the analogy to a superfluid
strictly breaks down.

Finally, we remark that there exists yet another exact mapping of the
dispersive hydrodynamic magnetization equations \eqref{eq:4},
\eqref{eq:5} (one-dimensional, $\alpha = 0$) to a continuum model of a
hard core Boson
gas~\cite{balakrishnan_particle-hole_2009,rubbo_quantum_2012}.

\section{Nucleation of nonlinear structures in ferromagnetic thin
  films}
\label{sec:nucl-nonl-struct}

Inspired by classical, incompressible fluids and BECs, we discuss the
diverse nonlinear structures that can be nucleated from an
impenetrable obstacle in a ferromagnetic thin film.  We numerically
confirm the main points of our discussion in Sec. \ref{sec:numerical-results}.

In the near vacuum regime, when the magnetodynamics limit to a BEC
(cf.~Sec.~\ref{sec:bose-einst-cond}), it is possible to qualitatively
predict the resulting
dynamics~\cite{Frisch1992,Nore1993,Carusotto2006,El2006,Gladush2007,Sasaki2010}. In
the subsonic regime, $\mathrm{M}<1$, quantized V-AV pairs are expected
to be nucleated for sufficiently large diameter obstacles. This occurs
because the obstacle gives rise to a local acceleration of the flow
and the fluid velocity develops a transverse component,
$u_y=\mathbf{u}\cdot\hat{\mathbf{y}}\neq0$. Locally, the total fluid
velocity reaches supersonic conditions, making the wake unsteady, periodically nucleating vortices. We stress that the global topology
of the magnetic texture in an infinite film must remain constant,
suggesting that only V-AV pairs can be nucleated. Consequently, a von
K\'{a}rm\'{a}n vortex street composed of single vortices of
alternating circulations is not favorable in ferromagnets although an
analog utilizing V-AV pairs has been numerically~\cite{Sasaki2010} and
experimentally~\cite{Kwon2016} observed in BEC.

As V-AV pairs are nucleated, two types of interactions are
possible. On the one hand, Vs and AVs are attractive and can
annihilate by transferring their energy to spin
waves~\cite{Hertel2006}. On the other hand, when there is an
underlying flow, a V-AV pair can form a stable entity exhibiting two
types of dynamical behavior, which we describe in a qualitative
fashion inspired by the well-known dynamics of vortices in classical
incompressible fluids~\cite{Leweke2016} and
BECs~\cite{Frisch1992,Nore1993}. If each V-AV pair is decoupled from
other V-AV pairs, the flow translates the V-AV pairs in a
train~\cite{Leweke2016}, similar to the Kelvin motion of same-polarity
V-AV pairs studied in planar ferromagnets with $\mathbf{u} =
0$~\cite{papanicolaou_semitopological_1999}. However, such a train of
parallel V-AVs was numerically observed to be unstable for propagation
according to the hydrodynamic equations \eqref{eq:12} for a
BEC~\cite{Nore1993} and may accommodate sinuous or varicose modes,
where a sinusoid describes the position of the V-AV pairs or the Vs
and AVs in the train, respectively. When the V-AV pairs are
sufficiently close to each other, V-V and AV-AV interactions can take
place, leading to their rotation about a vorticity
center~\cite{Leweke2016}. As we show in the next section, these
dynamics are observed numerically in magnetic hydrodynamic flow past
finite-sized obstacles.

In the supersonic regime, a distinguishable feature is a Mach cone
corresponding to steady, small amplitude, long waves. The aperture
angle of the Mach cone, referred to as the Mach angle $\mu$, can be
determined by assuming steady, small-amplitude, long-wavelength
perturbations nucleated from the obstacle (see, e.g.,
Ref.~\onlinecite{Courant1948}). Because these are two-dimensional
perturbations, we assume a wave vector
$\mathbf{k}=k_x\hat{\mathbf{x}}+k_y\hat{\mathbf{y}}$ with
$|\mathbf{k}|\ll1$. Let us also assume that the fluid and external
velocities have only $\hat{\mathbf{x}}$ components. The resulting
\textit{time-dependent} long-wavelength dispersion relation is just
Eq.~\eqref{eq:6} keeping only the linear in $|\mathbf{k}|$ terms,
\begin{equation}
  \label{eq:14}
  \omega_\pm = \left(2 \bar{n} \bar{u}+\bar{V}\right)k_x \pm
  \sqrt{k_x^2+k_y^2}\sqrt{(1-\bar{n}^2) (1-\bar{u}^2)}.
\end{equation}
The Mach cone is established as a steady state, thus we are interested in the relationship between $k_x$ and $k_y$ when $\omega_\pm=0$. 
The Mach angle $\mu$ can be calculated by trigonometric identities as
$\tan{\mu}=k_x/k_y$. Incorporating this transformation into
Eq.~\eqref{eq:14} with $\omega_\pm = 0$ and squaring leads to
\begin{equation}
  \label{eq:15}
  \left(2 \bar{n} \bar{u}+\bar{V}\right)^2\sin^2{\mu}= (1-\bar{n}^2)
  (1-\bar{u}^2).
\end{equation}
Solving Eq.~\eqref{eq:15} for the cases $\bar{V}=0$ and $\bar{u}=0$ leads to the definition of $\mu_u$ and $\mu_V$, respectively
\begin{subequations}
\begin{eqnarray}
\label{eq:161}
    \mu_u &=&
    \sin^{-1}{\frac{\sqrt{(1-\bar{n}^2)(1-\bar{u}^2)}}{2|\bar{n}\bar{u}|}}
    \le \frac{\pi}{2},\\   
\label{eq:162}
    \mu_V &=&
    \sin^{-1}{\frac{\sqrt{1-\bar{n}^2}}{\bar{V}}}=
    \sin^{-1}{\frac{1}{\mathrm{M}_V}} \le \frac{\pi}{2}.     
\end{eqnarray}
\end{subequations}
Note that Eq.~\eqref{eq:162} reduces to the Mach angle of a classical
gas with Mach number $\mathrm{M}_V$~\cite{Courant1948}. In contrast,
Eq.~\eqref{eq:161} is a more complex expression of $\bar{u}$ and
$\bar{n}$; note that $\mu_u$ and $\mu_V$ are real-valued only for
$\mathrm{M}_{u,V} \ge 1$ ($\mathrm{M}_{u,V} = 1$, $\mu_{u,V} = \pi/2$).
The non-standard form of the Mach angle $\mu_u$ is another consequence
of the broken Galilean invariance of the magnetic system.

\begin{figure}[t] 
  \centering \includegraphics[width=2.5in]{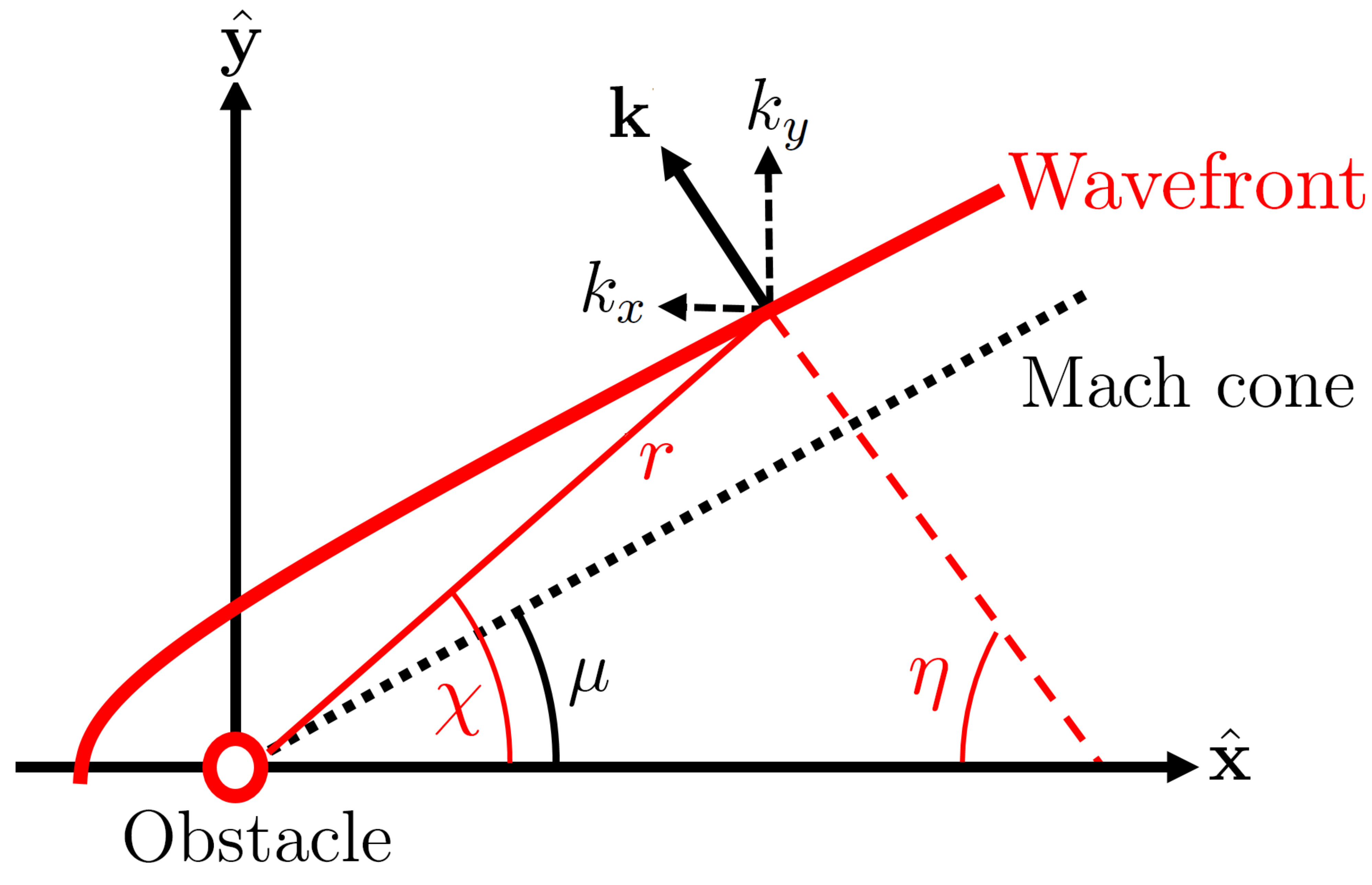}
  \caption{ \label{fig3} (color online) Schematic representation of a wavefront constant phase line (red solid line), traced by $\chi(\eta)$ and $r(\chi,\eta)$. The wavefronts are asymptotically parallel to the Mach cone (black dashed line) far from the obstacle (red circle). } 
\end{figure}

In steady flow, nonlinear structures are expected to reside inside the
Mach cone~\cite{El2006}. Outside the Mach cone, a static, $\omega=0$,
structure can also be established. This linear wavefront, also
referred to as Cerenkov radiation~\cite{Carusotto2006,Yan2013},
features a wave vector with constant phase curves describing an
approximate parabola around the obstacle. As outlined in
Ref.~\onlinecite{Gladush2007}, the constant phase curves can be
described in polar coordinates $(r,\chi)$ defined as a function of the
angle $\eta$, schematically shown in Fig.~\ref{fig3}. This is achieved
by assuming slowly modulated, stationary waves
$\mathbf{k}=|\mathbf{k}|\left(\cos\eta\hat{\mathbf{x}}+\sin\eta\hat{\mathbf{y}}\right)=\nabla\theta$. From
the static condition $\omega_\pm=0$ in Eq.~\eqref{eq:14}, it follows
that
\begin{equation}
\label{eq:173}
  |\mathbf{k}|^2 = \left(2\bar{n}\bar{u}-\bar{V}\right)^2\cos^2{\eta}-(1-\bar{n}^2)(1-\bar{u}^2).
\end{equation}

This implies, for constant $\eta$, that larger fluid or observer
velocities lead to shorter wavelengths. For time-independent
$\mathbf{k}$ and noting that $\nabla\times\mathbf{k}=0$ due to
irrotationality, it is possible to introduce a hyperbolic equation for
$k_x$ and $k_y$ that can be solved by the method of
characteristics~\cite{El2006}. Integrating $\nabla\theta$ along constant $\chi$ yields
\begin{subequations}
\label{eq:17}
\begin{eqnarray}
\label{eq:171}
  \tan\chi &=& -\frac{\partial\omega/\partial k_y}{\partial\omega/\partial k_x},\\
\label{eq:172}
  |\mathbf{r}| &=& -\frac{\theta}{|\mathbf{k}|\cos{\psi}},
\end{eqnarray}
\end{subequations}
where $\psi$ is the angle between $\mathbf{r}$ and $\mathbf{k}$. For
fixed phase $\theta$, Eqs.~\eqref{eq:171} and \eqref{eq:172} represent
the solution of a constant phase curve, as schematically represented
by the thick red solid line in Fig.~\ref{fig3}. It is also possible to recast this
solution in Cartesian components $(\mathrm{x},\mathrm{y})$. In the
moving reference frame with $\bar{u} = 0$, $\bar{V} \ne 0$, the result
coincides with that presented in Ref.~\onlinecite{Gladush2007}. In the
static reference frame $\bar{u} \ne 0$, $\bar{V} = 0$, we obtain by
trigonometric identities and algebra
\begin{subequations}
\label{eq:18}
\begin{eqnarray}
\label{eq:181}
\mathrm{x} &=&
\frac{\theta}{|\mathbf{k}|}\frac{\cos\chi}{\cos\psi}=\frac{\theta\cos\eta\left(\sin^2\mu_u+1-2\cos^2\eta\right)}{2\bar{n}\bar{u}\left(\cos^2\eta-\sin^2\mu_u\right)^{3/2}},\\ 
\label{eq:182}
  \mathrm{y} &=& \frac{\theta}{|\mathbf{k}|}\frac{\sin\chi}{\cos\psi}=-\frac{\theta\sin\eta\left(\cos^2\eta-[\sin^2\mu_u]/2\right)}{2\bar{n}\bar{u}\left(\cos^2\eta-\sin^2\mu_u\right)^{3/2}}.
\end{eqnarray}
\end{subequations}

It must be noted that far away from the obstacle, when $|\eta|\rightarrow\pi/2$ and in the limit of long-wavelengths $|\mathbf{k}|\ll1$, Eq.~\eqref{eq:171} approaches the solution $\chi\rightarrow\mu$. Therefore, the wavefronts are asymptotically parallel to the Mach cone. Close to the obstacle, when $|\eta|\ll1$, a series expansion yields the approximate parabolic wavefront profile
\begin{equation}
\label{eq:19}
  \mathrm{x}\sim-\theta+\mathrm{y}^2\frac{(2\bar{n}\bar{u})^2(1-\sin^2\mu_u)^2(4+\sin^2\mu_u)}{|\theta|(2-\sin^2\mu_u)}
\end{equation}

\section{Numerical results}
\label{sec:numerical-results}

The existence and dynamics of nonlinear structures hypothesized in
Sec.~\ref{sec:nucl-nonl-struct} for planar ferromagnetic thin films
can be verified by micromagnetic simulations. We consider impenetrable
obstacles with a circular cross section and diameter $d$. Due to
broken Galilean invariance, we perform simulations in both the static
and moving reference frames, as described below. It is important to
point out that the continuous nucleation of topological structures
requires an energy source, such as spin injection at a ferromagnetic
boundary~\cite{Sonin2010,Takei2014}. In our simulations, we do not
consider such an energy source per se. Rather, the stabilization of
nonzero velocity $\mathbf{u}$ [although static, $-(1-n^2)\mathbf{u}$ is
nevertheless a spin current~\cite{Iacocca2016b}] is assumed to result
from some external mechanism, analogous to the sustenance of a flowing
classical fluid. The introduction of an obstacle is a perturbation to
the system that imposes a new energy minimum. This leads to a
transient regime where topological structures are nucleated as the
total energy of the system is minimized. Therefore, we report and
interpret the dynamics observed during such a transient from a
dispersive hydrodynamic perspective.

\subsection{Static reference frame, $\bar{u}\neq0$ and $\bar{V}=0$}

We perform finite-difference integration of the LL equation using the
GPU-accelerated micromagnetic package
MuMax3~\cite{Vansteenkiste2014}. Although we report solutions of the
nondimensional LL equation, Eq.~\eqref{eq:1}, the parameters are
consistent with Permalloy, e.g., saturation magnetization
$M_s=790$~kA/m, exchange stiffness $A=10$~pJ/m$^3$, and Gilbert
damping $\alpha=0.01$. To compare with theory, we consider only local
dipolar fields (zero thickness limit) by incorporating a negative
perpendicular anisotropy constant $K_u=-\mu_0M_s^2/2$, where $\mu_0$
is the vacuum permeability. With these parameters, space is normalized
to the exchange length $\lambda_\mathrm{ex}$ and time to
$1/(\gamma\mu_0M_s)$, where $\gamma$ is the gyromagnetic ratio.
Typical values for these parameters are $\lambda_\mathrm{ex} = 5$~nm,
$1/(\gamma \mu_0 M_s) = 36$~ps, and $\gamma =28$~GHz/T.

An impenetrable obstacle is introduced as a localized
(hyper-Gaussian), perpendicular field that forces $\bar{n}=1$, i.e.,
the vacuum state, in a limited area~\cite{Iacocca2016b}. In fact, a
perpendicular field gradient acts as a potential body force on the fluid as
shown in Eq.~\eqref{eq:52}. Alternatively, a magnetic defect or
absence of magnetization also constitutes an impenetrable
object. As an initial condition, we impose
a SDW given by $(\bar{n},\bar{u}=0.6)$ with $0 < \bar{n} < 1$, and we guarantee its stability both by applying a homogeneous, perpendicular
magnetic field that satisfies Eq.~\eqref{eq:53} and by defining a
simulation domain of $L_x\times L_y=317\times156$ spatial units that accommodates an exact number
of SDW periods in the $\hat{\mathbf{x}}$ direction. The simulation is
discretized into a mesh with $1024\times512$ gridpoints and we implement periodic
boundary conditions along the $\hat{\mathbf{x}}$ direction and open or
free-spin conditions in the $\hat{\mathbf{y}}$ direction. The
simulation is evolved to the time $t=112$. In order to prevent any
nucleated structure from propagating through the periodic boundaries,
we also implement a high damping region close to the edge of the
simulation domain. Such an absorbing boundary does not compromise the
stability of the initialized SDW. We emphasize that the choice of
fluid velocity $\bar{u}=0.6$ was made on the basis of computational
speed.  Other fluid velocities give similar results.
\begin{figure}[t!]
\centering \includegraphics[width=3.in]{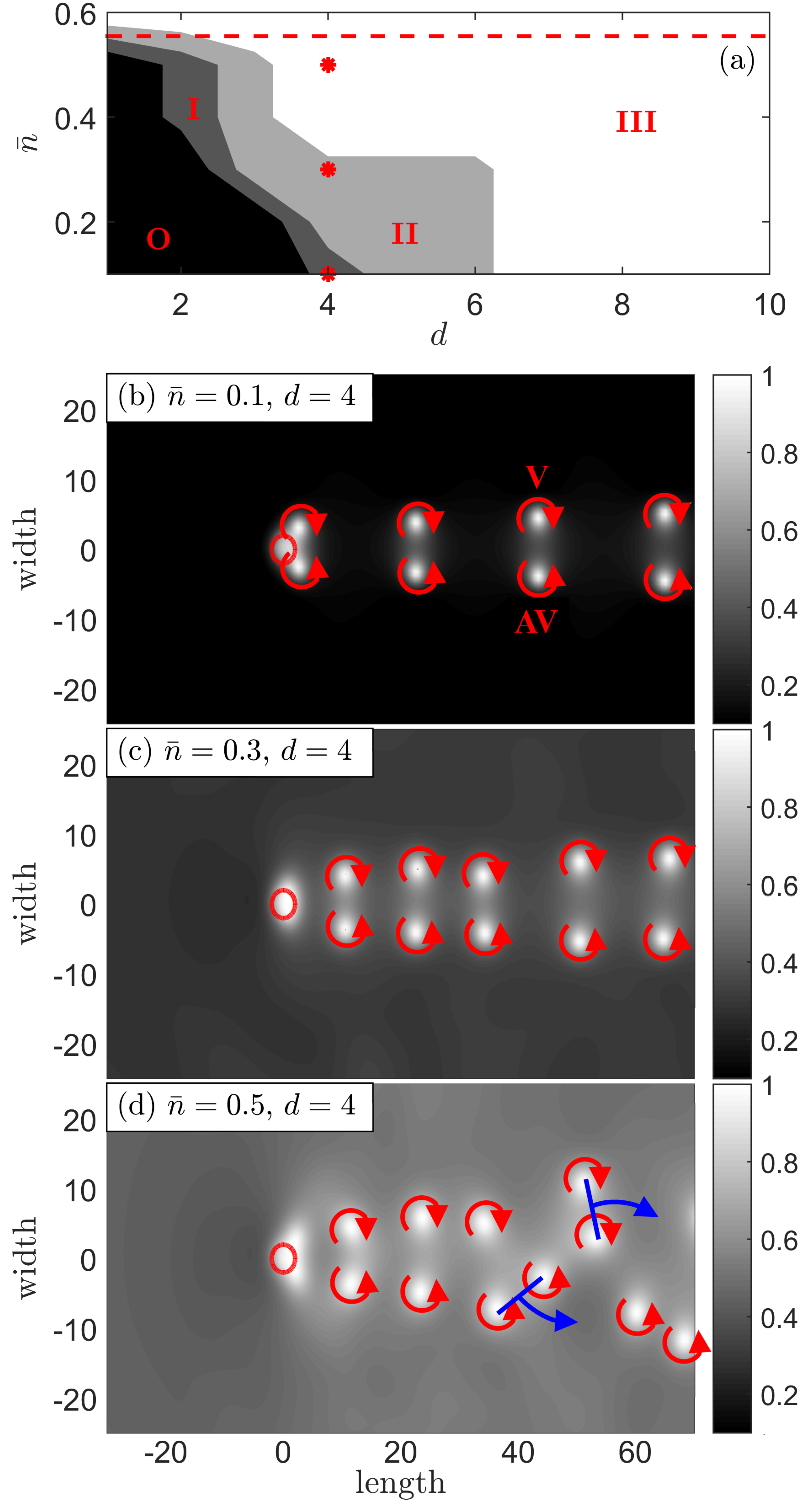}
\caption{ \label{fig4} (color online) (a) Phase diagram for V-AV
  nucleation and dynamics in the static reference frame. Region O
  represents laminar flow while the gray regions I-II (III) denote
  ordered (irregular) V-AV pair nucleation. Snapshots of the
  longitudinal spin density, $n$, exemplify each region showing V-AV
  pairs in (b) parallel (region I), (c) unstable mode (region II), and
  (d) irregular dynamics (region III), with corresponding conditions
  indicated by red asterisks in (a). The red arrows represent the
  vortices' circulation direction, with V and AV circulations
  indicated in (b). The blue lines and arrows schematically show the
  rotating trajectory of the V-V and AV-AV pairs in the irregular
  regime. }
\end{figure}

The observed nucleated nonlinear structures can be classified in the
phase space spanned by the parameters $\bar{n}$ and $d$, shown in
Fig.~\ref{fig4}(a), swept in steps of $0.1$ and $1$, respectively. First, we note that laminar flow (region O) is lost as
the diameter of the obstacle increases, even in the subsonic regime
(below the dashed line), indicating that the fluid velocity locally
develops supersonic speed as it is accelerated around the obstacle. We
have verified this fact numerically. The gray regions represent
ordered V-AV pair nucleation. For a relatively narrow parameter space
(region I), the V-AV pairs translate parallel to each other carried by
the underlying flow, as shown in Fig.~\ref{fig4}(b). In region II, the parallel V-AV trains are unstable~\cite{Nore1993} [Fig.~\ref{fig4}(c)]. We note that
the instability numerically observed for parallel V-AV pairs in
BECs~\cite{Nore1993} suggests that region I could eventually develop
an instability for larger simulation domains and times. However, simulations in a domain twice as large $634\times212$, evolved twice as
long to $t=224$ did not show evidence of such an instability. Finally,
region III denotes irregular V-AV nucleation, as shown in
Fig.~\ref{fig4}(d). Here, V-V and AV-AV rotation is observed,
schematically shown in Fig.~\ref{fig4}(d) by the blue lines defining
the V-V and AV-AV axis and blue arrows denoting the rotation. In all the cases mentioned above, we observe that the separation between the V and AV in a pair increases as they translate away from the obstacle, similar to a Magnus force. This was also observed in additional simulations in which a single V-AV pair was nucleated by a field pulse. Such a single pair moves with a velocity greater than $|\mathbf{u}|$ and develops a velocity transverse to $\mathbf{u}$. This is in contrast to constant V-AV motion in a uniform magnetic background~\cite{papanicolaou_semitopological_1999}. An in-depth study of V-AV motion on a textured background is worthy of future investigation, but it lies outside the scope of the present paper.

\begin{figure}[t!]
\centering \includegraphics[width=3.in]{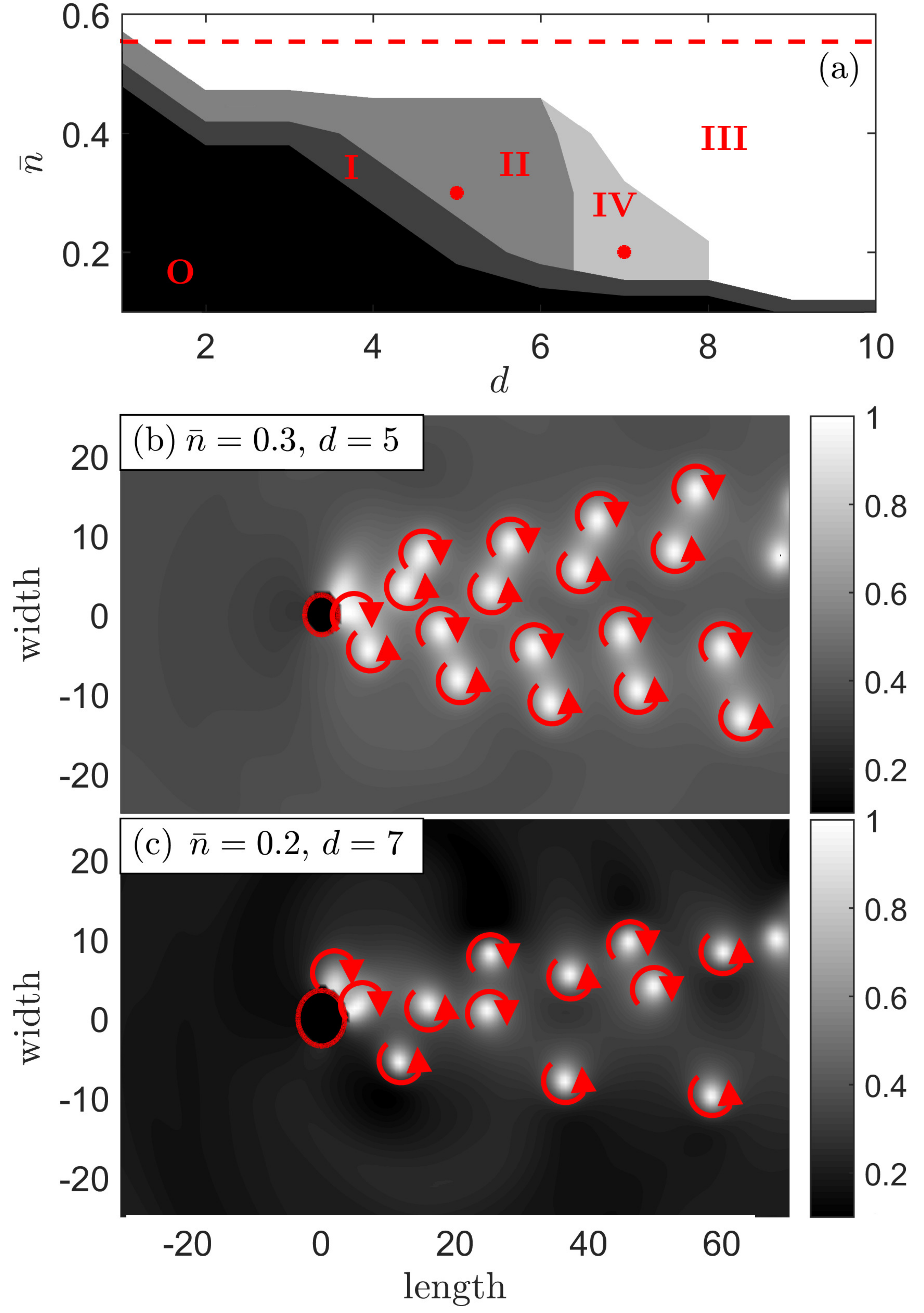}
\caption{ \label{figapp1} (color online) (a) Phase diagram for V-AV
  nucleation and dynamics in the static reference frame considering a magnetic defect as an obstacle. The black region
  represents laminar flow while the gray regions (white region)
  denotes ordered (irregular) V-AV pairs nucleation. Regions I and II
  qualitatively agree with the dynamics observed in
  Fig.~\ref{fig4}. Here, region II exhibits a sinuous mode (b), in
  contrast to Fig.~\ref{fig4}. Additionally, a von K\'{a}rm\'{a}n-like
  vortex street is observed in region IV (c). We do not show regions I
  and III as they are similar to Figs.~\ref{fig4}b and \ref{fig4}c,
  respectively.}
\end{figure}

Micromagnetic simulations were performed also in the case in which a magnetic void with free spin boundary conditions serves as an obstacle. The resulting $\bar{n}$ vs $d$ phase space is shown in Fig.~\ref{figapp1}(a) and agrees qualitatively with the phase diagram obtained with a localized field [Fig.~\ref{fig4}(a)]. We note that in this case, the instability of the V-AV pairs train develops
into a sinous mode [Fig.~\ref{figapp1}(b)]. Furthermore, we observe a
transition between region II and III, labeled region IV in
Fig.~\ref{figapp1}(a), where we observe V-V and AV-AV nucleation reminiscent of von K\'{a}rm\'{a}n vortex streets numerically predicted and recently observed in BECs~\cite{Sasaki2010,Kwon2016}.

\begin{figure}[t]
\centering \includegraphics[width=2.8in]{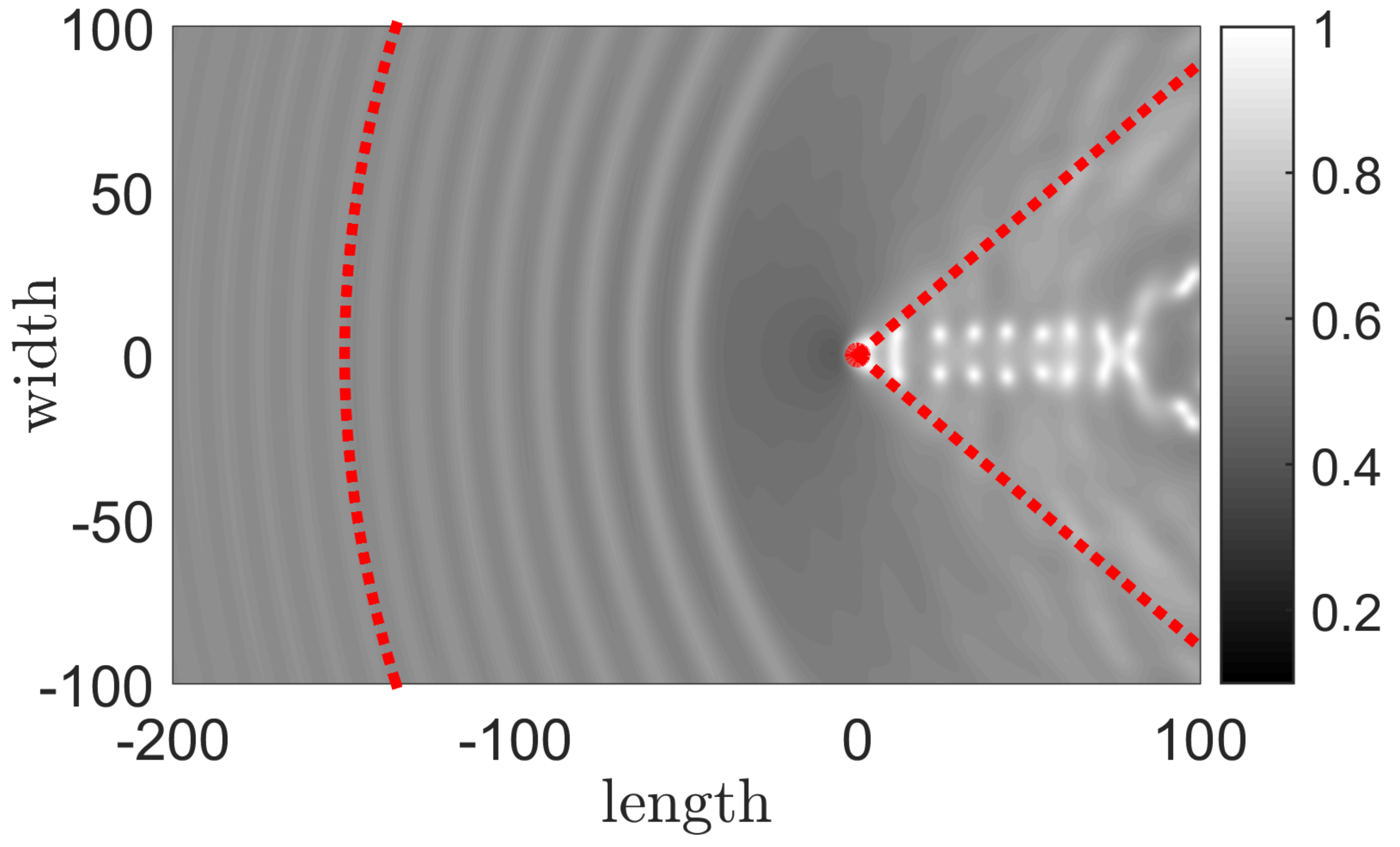}
\caption{ \label{fig6} (color online) Snapshot of the longitudinal
  spin density, $n$, in the static reference frame at a supersonic
  condition $(\bar{n},\bar{u})=(0.6,0.6)$ and an obstacle of diameter
  $d=6$. The Mach cone calculated from Eq.~\eqref{eq:161} and the
  wavefront calculated from Eq.~\eqref{eq:18} are shown by red dashed
  curves.}
\end{figure}

Irregular V-AV nucleation persists for supersonic conditions, when
$M_u>1$ [above the red dashed line in
Fig.~\ref{fig4}(a)]. Additionally, in this regime we also observe a
well-defined Mach cone and the nucleation of wavefronts (Fig.~\ref{fig6}). As discussed in Sec.~\ref{sec:nucl-nonl-struct}, the
Mach angle and curves of constant phase corresponding to wavefronts
are given in Eqs.~\eqref{eq:161} and \eqref{eq:18}, which describe the
numerical results to good accuracy, as shown by red dotted lines in
Fig.~\ref{fig6}. Whereas the Mach angle defines a static boundary
i.e., the Mach cone, the wavefronts are dynamic due to vortex shedding and the concomitant change of flow conditions effected by our energy minimizing simulations. For this reason, the estimates of Eq.~\eqref{eq:173} are only valid for timescales shorter than the relaxation time imposed by damping. By fitting $\theta$ and a horizontal shift from the obstacle, the red dashed line outlining the wavefront in Fig.~\ref{fig6} is obtained. The wavelength
$2\pi/|\mathbf{k}|=\lambda=19$ at $\eta=0$ is well-described by
Eq.~\eqref{eq:173} and is found to be within the same order of
magnitude as the numerically calculated wavefront wavelengths
throughout the simulation evolution
$\lambda_\mathrm{sim}=12.21\pm2.1$. This agreement is possible due to
the fact that the equations are effectively
conservative for short timescales and the concepts used for BEC are approximately
applicable.

\subsection{Moving reference frame, $\bar{u}=0$ and $\bar{V}\neq0$}

In the moving reference frame, we use a pseudo-spectral
method~\cite{Hoefer2012b} to solve the nondimensionalized LL equation,
Eq.~\eqref{eq:1}. Here, the localized, perpendicular field moves with
velocity $\bar{V}$ while the fluid velocity is zero. The simulation
domain in this case has the same size as the simulations in the static
reference frame but it is discretized into the coarser $256\times128$
number of grid points by virtue of the accuracy of the pseudo-spectral
method. We initialize the simulation with a homogeneous magnetization
with $\bar{n}=h_0=0.7$, $\bar{u}=0$ and we use periodic boundary
conditions along and across the thin film.  We evolve the simulation
to $t=2000$. The longer simulation time in this case reflects the
slower dynamics in the moving reference frame. The value of the
homogeneous perpendicular field bias magnitude, $h_0$, was chosen to
minimize the strength of the localized perpendicular field needed to
impose hydrodynamic vacuum.

\begin{figure}[t]
\centering \includegraphics[width=3.in]{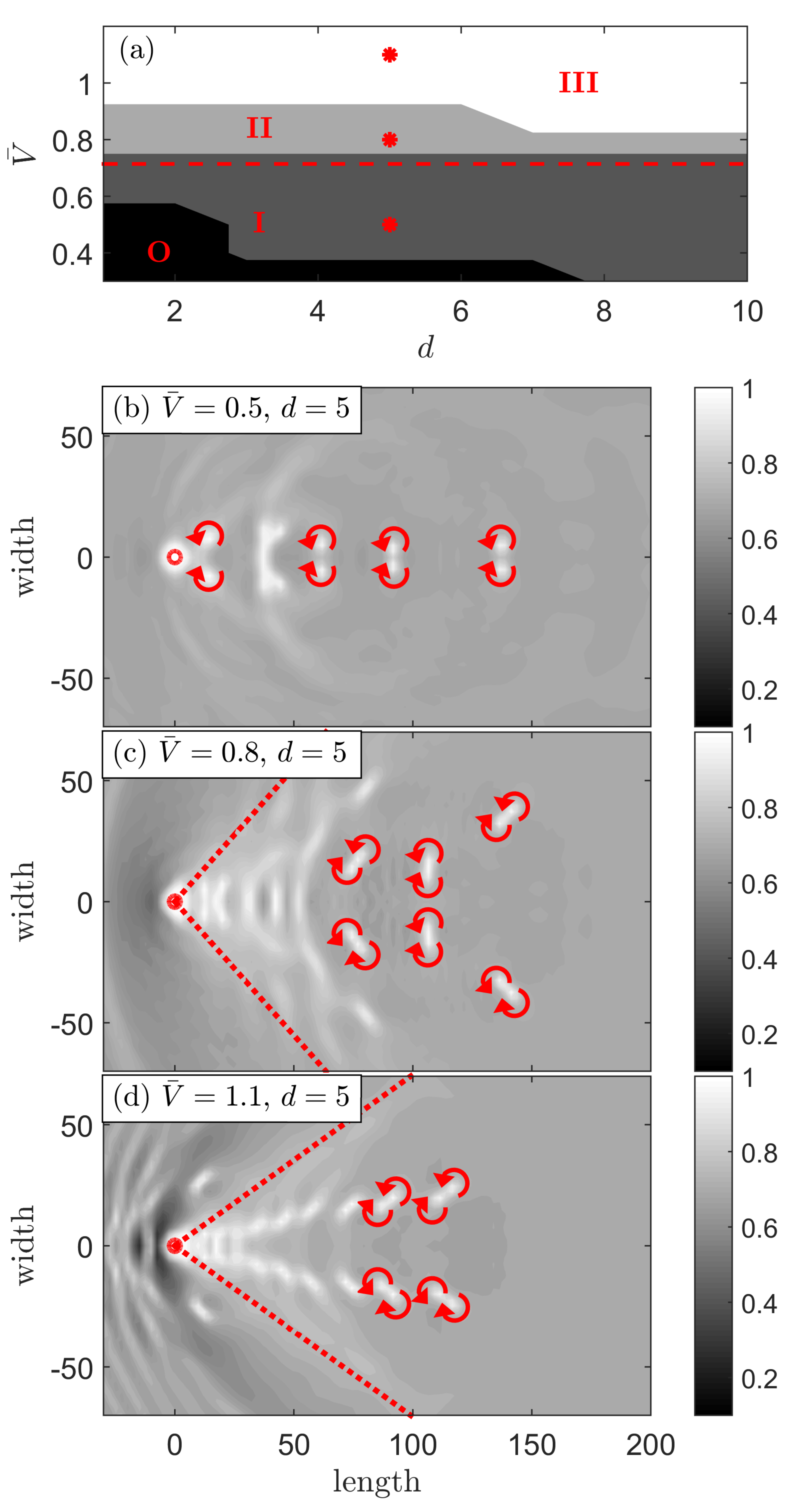}
\caption{ \label{fig8} (color online) (a) Phase diagram for V-AV
  nucleation and dynamics in the moving reference frame. Region O
  represents laminar flow, where vortices are not nucleated. The gray
  regions I-II (region III) denote irregular (ordered) V-AV pair
  nucleation. Snapshots of the longitudinal spin density, $n$, show
  V-AV (b) annihilation (region I), (c) irregular dynamics (region II)
  and (d) ordered dynamics (region III) inside the Mach cone with
  conditions indicated by red asterisks in (a). The red arrows
  represent the vortices' circulation direction. }
\end{figure}

The phase space for the nucleated nonlinear structures and dynamics as
a function of $\bar{V}$ and $d$ are shown in Fig.~\ref{fig8}(a), swept in steps of $0.1$ and $1$, respectively. As in
Fig.~\ref{fig4}(a), subsonic laminar flow (region O) is lost as the
obstacle diameter increases. Regions I to III denote V-AV pair
nucleation. In contrast to the static reference frame, the absence of an underlying fluid velocity in this case, $\bar{u}=0$, precludes significant vortex translation or Kelvin motion. Instead, the V-AV
pairs are attracted to each other due to magnetic damping (energy
dissipation) and subsequently annihilate, expelling spin waves. For
this reason, the V-AV pairs nucleated at subsonic conditions (region
I) are unstable and lead to an unevenly spaced V-AV train as well as
irregular dynamics close to the obstacle [Fig.~\ref{fig8}(b)]. At
supersonic conditions, above the red dashed line in
Fig.~\ref{fig8}(a), a narrow phase space (region II) leads to
irregular V-AV pairs inside the Mach cone, shown in Fig.~\ref{fig8}(c)
by red dashed lines calculated from Eq.~\eqref{eq:162}. Finally, in
region III, the V-AV pairs become mostly ordered. In
Fig.~\ref{fig8}(d) we observe two V-AV pairs establishing an oblique
path with respect to the obstacle. This is reminiscent of oblique
solitons~\cite{El2006}, but here the structure immediately breaks down
into V-AV pairs.

\begin{figure}[t]
\centering \includegraphics[width=3.in]{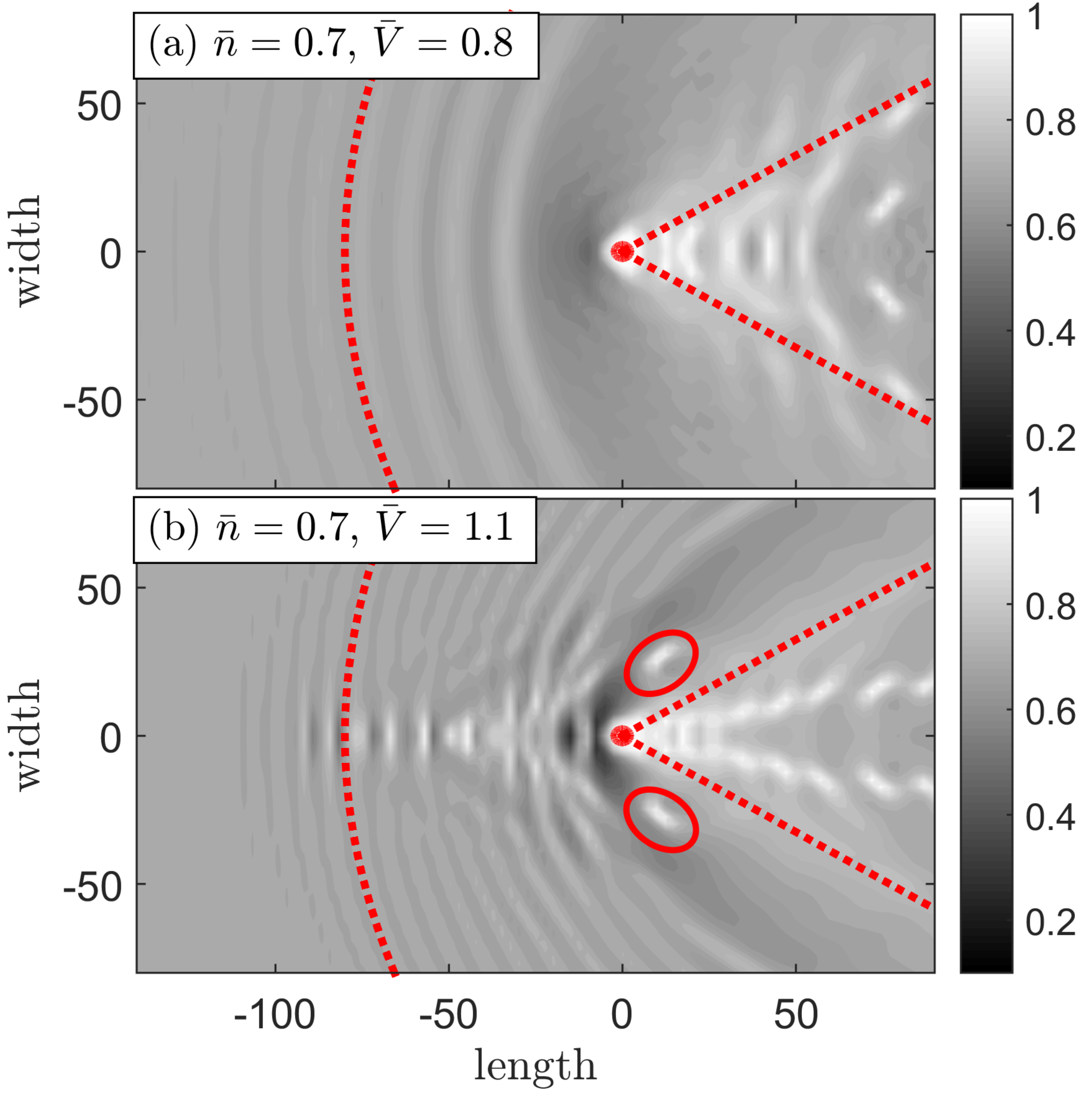}
\caption{ \label{fig9} (color online) Snapshots of the longitudinal
  spin density, $n$, in the moving reference frame at supersonic
  conditions. The moving localized field has a diameter $d=5$. The
  Mach cone calculated from Eq.~\eqref{eq:161} and the wavefront
  calculated from Eq.~\eqref{eq:171} are shown by red dashed
  lines. Instabilities in the wavefronts lead to V-AV pairs, red
  circles. }
\end{figure}
As discussed above, a Mach cone and wavefronts are also established at
supersonic conditions in the moving reference frame. These agree to
good accuracy with Eqs.~\eqref{eq:162} and \eqref{eq:17}, where the
curves of constant phase for $\rm x$ and $\rm y$, analogous to
Eq.~\eqref{eq:18}, are determined as in \onlinecite{Gladush2007}.
Figure~\ref{fig9} shows the comparison between theory and
numerics. However, we observe instabilities in the wavefronts that
develop for large obstacle velocities. This is shown in
Fig.~\ref{fig9}(a) and (b) for, respectively, velocities similar to and
larger than the sonic curve. In particular, the short wavefronts in
Fig.~\ref{fig9}(b) are observed to break into V-AV pairs due to this
instability (encircled in red) suggesting nonlinear effects. The study
of such effects is beyond the scope of the current paper.

\section{Conclusion}

In summary, we demonstrate that nonlinear structures and V-AV
complexes can be nucleated from an obstacle in a thin film planar
ferromagnet. These observations are in qualitative agreement with
structures nucleated in classical fluids and superfluids, providing further
evidence for the hydrodynamic properties of thin film planar
ferromagnets.

Both in the moving and static reference frames, we observe nucleation
of V-AV pairs, as qualitatively expected for an unsteady wake formed
behind an impenetrable obstacle. These V-AV pairs experience diverse
dynamics and are nucleated as long as the system is out of
equilibrium. In the static reference frame, we observe Kelvin motion,
instability, and V-V, AV-AV rotation. In the moving reference frame,
V-AV pairs are also nucleated but generally annihilate, forming an
irregular pattern and spin waves. For supersonic conditions, a Mach
cone and the formation of wavefronts are observed, as expected for
superfluids. Good quantitative agreement is found between the
numerically observed structures and theoretical results derived from
the linearized, conservative, long-wave equations.

Although nonlinear structures and V-AV complexes are observed in both
the static and moving reference frames, there are important
differences in their dynamics. In particular, irregular V-AV dynamics
are observed in supersonic conditions in the static reference frame,
$\bar{u}>0$ and $\bar{V}=0$, and subsonic conditions in the moving
reference frame, $\bar{u}=0$ and $\bar{V}>0$. This is due to the fact
that the underlying flow, $\bar{u}\neq0$, induces V-AV translation in
contrast to the moving reference frame with $\bar{u}=0$. In other
words, a topological texture is required to support ordered vortex
structures. This suggests that for a fixed $\bar{u}>0$ the
introduction of $\bar{V}\neq0$ leads to irregularity in the subsonic
regime.  Additionally, the numerical observation of an apparently
stable V-AV train propagating on the textured background $\bar{u} > 0$
is to be contrasted with the unstable propagation of a train of
counter-rotating vortices in a BEC.  This intriguing ordered regime
requires further analysis.

It is noteworthy that we explore a two-dimensional parameter space
with a fixed $\bar{u}$ in the static reference frame and a fixed
$\bar{n}$ in the moving reference frame, as opposed to the full
three-dimensional parameter space where $\bar{n}$, $d$, and $\bar{u}$
or $\bar{V}$ are varied. However, we argue that the phase spaces shown
in Fig.~\ref{fig4}(a) and Fig.~\ref{fig8}(a) represent general
qualitative features of the full parameter space. The phase space can
collapse to two dimensions, spanning $\mathrm{M}_u =
\mathrm{M}_u(\bar{n},\bar{u})$ or $\mathrm{M}_V =
\mathrm{M}_V(\bar{n},\bar{V})$ versus $d$. Micromagnetic simulations
performed with several choices of $\bar{n}$ and $\bar{u}$ and
$\bar{V}$ for, respectively, the static and moving reference frames,
indeed return qualitatively similar dynamics for matching Mach numbers
and obstacle diameters.

The above results show that V-AV complexes can be nucleated in a thin
film ferromagnet with planar anisotropy following well-defined
patterns analogous to superfluids and compressible fluids. Our observations
are relevant for the stability of spin-density waves and the study of
V-AV interactions with other nonlinear textures such as spin-density
waves, other V-AV pairs, and wavefronts.

\begin{acknowledgments}
  E.I. acknowledges support from the Swedish Research Council,
  Reg. No. 637-2014-6863. M.A.H. partially supported by NSF CAREER
  DMS-1255422.
\end{acknowledgments}

\bibliographystyle{aipnum4-1}

\end{document}